Hybrid density functional theory study on zinc blende GaN and diamond surfaces and interfaces: Effects of size, hydrogen passivation and dipole corrections

Eric Welch[a,*] and Luisa Scolfaro[a]

[a]Department of Physics, Texas State University, San Marcos, Texas, United States

Abstract

GaN based high electron mobility transistors show promise in numerous device applications which elicits the need for accurate models of bulk, surface, and interface electronic properties. We detail here a hybrid density functional theory study of zinc blende (zb) GaN and diamond bulk and surface properties, and zb GaN on diamond interfaces using slab supercell models. Details are provided on the dependence of electronic properties with respect to supercell size, the use of pseudo-hydrogen to passivate the bottom GaN layer, and dipole corrections. The large bulk modulus of diamond provides a templating structure for GaN to grow upon, where a large lattice mismatch is accounted for through the inclusion of a cationic Ga adlayer. Looking at both type I and II surfaces and interfaces of GaN shows the instability of zb GaN without an adlayer (type II), where increased size, pseudo-hydrogen passivation and dipole corrections do not remove the spurious interaction between the top and bottom layers in type II GaN. Layer dependent density of states, local potential differences, and charge density differences show that the type I interface (with a Ga adlayer) is stable with an adhesion energy of 0.704 eV/Å$^2$ (4.346 J/m$^2$); interestingly, the diamond charge density intercalates into the first layer of GaN, which was seen experimentally for wurtzite GaN grown over diamond. The type II interface is shown to be unstable which implies that, to form a stable, thin-film zb interface between GaN and diamond, the partial pressure of trimethylgallium must be controlled to ensure a Ga layer exists both on the top and bottom layer of the GaN thin film atop the diamond. We believe our results can shed light towards a better understanding of the GaN/diamond multifaceted interface present in the GaN overgrowth on diamond samples.

**1. Introduction**

High electron mobility transistors (HEMTs) using gallium nitride (GaN) have been studied extensively since their inception. Compared to standard HEMT channel materials like Si, GaN exhibits a wider band gap, a larger two dimensional electron gas (2DEG) region, larger electron saturation velocity at high frequencies and a higher breakdown voltage[1]. Applications for GaN HEMTs include microwave satellite[2], wireless and radio communications devices[3], monolithic microwave integrated circuits (MMICs)[4,5], power amplifiers[6–8], and power switches for high voltage devices[9]. To realize high efficiency GaN HEMTs, however, high quality heterostructures must be designed around the GaN channel layer.

Substrate options in early research into GaN growth included SiC, GaAs, ZnO, AlN, native GaN, and sapphire ($Al_2O_3$), where sapphire was determined to be the lowest cost, highest performance option due to increased power gain, a better lattice mismatch, and improved semi-insulating properties[3]. More recent studies have shown that due to enhancement of thermal transport, however, chemical vapor deposition (CVD) grown diamond is a more viable substrate option; diamond has properties similar to SiC, but with higher thermal conductivity[10,11]. The growth of this heterostructure (GaN-on-diamond) has been developed and improved, and a recent report detailed the promising method of epitaxial lateral overgrowth to achieve high quality devices[11]. Thus, a full understanding of the interface between GaN polymorphs and diamond is desirable to help further understanding of device mechanisms and better control of their efficiency.

It is known that zb GaN and diamond are polar surfaces using Tasker's model, implying surface charge density redistribution is required for chemical stability[12]; with GaN, type III polar surfaces can exist under certain growth conditions which result in surface reconstruction or charge redistribution. During growth, alternating anionic and cationic layers form in zb GaN, while in both zb GaN and diamond, local inversion symmetry is broken due to different staking layer distances. To account for the polarity, pseudo-hydrogen passivation of the bottom layer of each supercell is used to remove spurious interactions between periodic images in the direction perpendicular to the surface. Numerous studies have shown the efficacy of pseudo-hydrogen passivation both for polar surfaces[13–20] and interfaces[21–23]. Also, to counteract the electric field created due to local non-centrosymmetry, dipole corrections are added to the energy and the potential of the system. With these corrections, an accurate view of the surface and interface electronic structures is possible.

There exist two types of interface structures for zb-zb heterostructures (and thus two types of surfaces in GaN), where the top and bottom layers of GaN are either the same (type I) or different (type II) atomically, as is shown in Figure 1. A large enough supercell would allow one to neglect these differences, but this study focuses on thin films and thus this difference must be considered. Both types of surfaces are explored in this study along with the constituent surface structures in GaN. It should be noted that the type I surface is effectively an adlayer surface reconstruction, where a single monolayer of the atoms comprising the bottom of the supercell are added to the top layer of the supercell. The addition of a Ga adlayer, along with H passivation has been shown to lower surface energy in wurtzite GaN[24] while cation excess leads to reduced stress during growth due to the metallization by a cation adlayer in zb GaN[20].

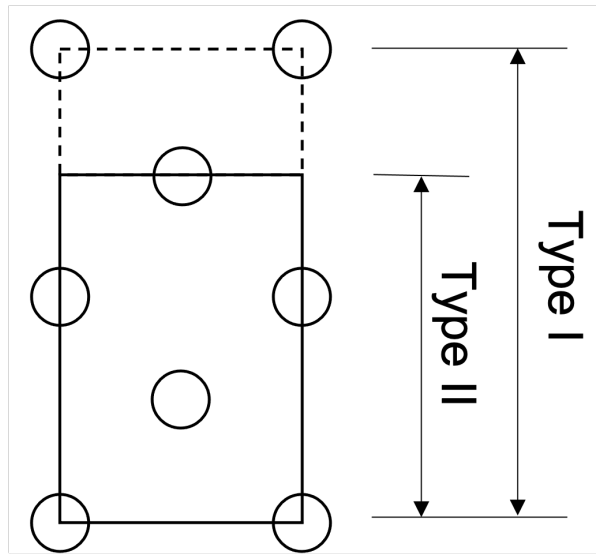

Figure 1. Toy model of type I and II polar surfaces. The dashed outer line encompasses the type I structure, while the solid line represents the type II structure.

Some studies on GaN surfaces utilize a wedge-shaped supercell to deal with these polar structures[24,25]. This however, is done for the wurtzite structure and often requires a very large supercell and vacuum region, while others have shown that slab models may also be used to study the electronic properties of GaN[18,26–28]; we employ the slab model throughout.

Recent experimental results have shown both zb and wurtzite (wz) polymorphs of GaN grown on top of rough diamond surfaces[29]; the existence of both GaN polymorphs is expected, as the formation energy of both structures are similar[30]. While there are numerous experimental[30–32] and computational[21,23,33–37] studies on wz GaN surfaces and non-diamond interfaces, diamond surfaces[38] and on computational modeling of polar wz surfaces and interfaces[18,19,24–28,39–41], little is known about the interface structure, bonding, energetics, and charge density distribution of GaN on diamond heterostructures. Thus, we show here hybrid density functional theory (DFT) calculations of bulk, surface, and interface properties of zb GaN and diamond to show the effects of supercell size, H passivation and dipole corrections on electronic properties. We choose zb GaN and its interface with diamond which represents a cubic/cubic structure for the present study.

This paper is organized as follows: Section 2 outlines the computational methods employed in this study; section 3 details the results of these calculations showing total and surface project density of states (DOS), surface-normal projected local potentials (LPOT), charge density differences (CDDs), and band edge projected charge densities; section 4 discusses conclusions drawn and future work building off these results. Note that results in the main paper are for (111) diamond, type I (111) GaN, and type II (111) GaN, while type I and type II (11$\bar{1}$) GaN results are shown in the supplemental information. The Ga terminated surfaces have been shown to interface with diamond in experiment, where intercalation of C into GaN is seen[29]. This is the interface of interest in these calculations, but all surface terminations were studied for consistency.

2. **Computational methods**

DFT total energy and electronic structure calculations were done using the Vienna Ab-Initio Simulation Package (VASP)[42–45] . The projector augmented wave (PAW) method[46] was implemented with Perdew-Burke-Ernzerhof (PBE) pseudopotentials. Local PBE revised for solids (PBEsol)[47] and non-local hybrid[48] were used for exchange-correlation (XC) within the generalized gradient approximation (GGA)[49]. C $2s^22p^2$, Ga $3d^{10}4s^24p^1$, and N $2s^22p^3$ electrons were treated as valence states while planewaves were expanded in a basis set truncated at 650 eV; $H_{0.75}$, $H_{1.25}$ or $H_{1.0}$ pseudo-hydrogen was used for N, Ga, or C passivated surfaces, respectively. Bulk structures of zb GaN and diamond were relaxed using the conjugate gradient method with electronic convergence of $10^{-5}$ eV in the total energy. 8x8x8 gamma centered k-point meshes were used for both bulk structures with the tetrahedron method with Blöchl correction used to account for partial occupancies.

Surface structures were built from equilibrium bulk structures for one- and three-unit cell supercells to elucidate size-based effects. The bottom three layers of each surface supercell were fixed at their bulk positions during relaxation, allowing the upper layers to relax until forces on ions dropped below 0.01 $eV$/Å. A vacuum region equal to the number of unit cells above the non-passivated surface was used in each surface model e.g., three-unit cell structures have three-unit cells worth of vacuum between periodic images in the surface normal direction with a total volume equal to three-unit cells of volume. For interfaces, constrained relaxation was done on three-unit cell interface supercells where selective dynamics was used to limit ionic relaxation along the *z*-axis, fixing the top and bottom three layers of the heterostructure at their bulk positions; a modified compilation of VASP is required for constrained relaxation. Passivation was also done on interface supercells at the upper and lower surfaces, where H atoms were fixed 1.5 Å away from the surface. 20 Å of vacuum was included on either side of the interface supercell (for a total of 40 Å of the supercell) to minimize H-H interactions between periodic supercell images.

Dipole corrections included in the total energy and potential are of the form

$$E_{dip} = \frac{1}{2}\int_\Omega [\rho^{ion}(\mathbf{r}) - \rho^{ele}(\mathbf{r})]V_{dip}(\mathbf{r})d^3r, \quad (1)$$

and,

$$V_{dip}(z) = 4\pi m\left(\frac{z}{z_m} - \frac{1}{2}\right), 0 < z < z_m, \quad (2)$$

where $\rho^{ion}$ ($\rho^{ele}$) is the ionic (electronic) charge density, $m$ is the surface dipole density of the slab and $z_m$ is the height of the supercell, and the integral is over the volume $\Omega$. Due to the asymmetric nature of the polar surface supercells, the surface dipole density is non-zero and the potential varies across the vacuum region. This variation in the potential results in an artificial dipole field at the surface which can be compensated for by including Eqns. (1) and (2) into the model. Consequently, these corrections result in a jump discontinuity in the potential which is centered in the vacuum region[50]. The corrections were added along the direction normal to the surface and done sequentially, first adding a dipole correction to the total energy in the direction normal to the surface and then to the total potential and forces of the energy-dipole corrected wavefunctions.

Along with anomalous dipole fields, interactions between images due to the boundary conditions used in DFT cause specious interactions between periodic supercell surfaces. To account for this interaction, a large vacuum region was used along with pseudo-hydrogen to passivate the bottom surface dangling bonds, both neutralizing charge on the bottom surface and stabilizing the layers which are held at their bulk positions during surface relaxation. The use of pseudo-hydrogen helps to remove erroneous electronic band gap states arising from interactions between periodic surfaces, allowing for the calculation of accurate surface energies and electronic properties[13].

The effects of dipole corrections and H passivation are quantified by projecting the potential in the slab along the direction normal to the surface (z-axis direction). The surface-normal projected local potential (LPOT) is given by

$$V_{POT}(z) = V(z) + \int \frac{\rho(z')}{|z-z|} dz', \tag{3}$$

where $V(z)$ is the potential due to the ions in the supercell and the integral is the Hartree potential. The Hartree term is a Coulomb-like potential in the Hartree-Fock theory due to the electronic charge density[50].

3. Results and Discussion

3.1. Bulk properties

Equilibrium lattice constants, bulk moduli and electronic band gaps are summarized in Table 1 for diamond and zb GaN. Equilibrium lattice constants were determined using local PBEsol exchange-correlation, which has been shown to produce some of the lowest errors in cohesive properties of semiconductor materials[51]; errors in the lattice constant were < 1% related to experimental values. This data was fit to the third order Birch-Murnaghan equation of state to obtain bulk moduli[52,53]. Electronic band gaps were obtained including hybrid non-local exchange-correlation (PBE0), with range separation included for diamond (HSE06), which resulted in errors of < 1%. Figure 1 shows the equilibrium bulk structure for diamond and zb GaN projected along the (111) direction, the surfaces of interest in this study. Plots of the bulk modulus data are in the supplemental information. Slight deviations from experimental results are due in part to the computationally limited basis-set and k-point mesh size as well as the inefficiencies of current exchange-correlation functionals, even at the hybrid functional level.

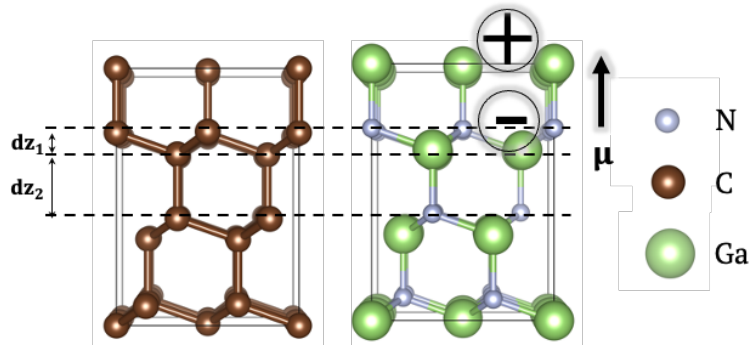

Figure 2. Structures of (111) diamond and (111) GaN where $dz_i$ is the layer spacing, $\mu$ is the electronic dipole moment and + (-) is a cation (anion) layer.

| System | $a_0$ (Å) | $B_0$ (GPa) | $E_g$ (eV) |
|---|---|---|---|
| diamond | 3.57 (3.567)[54] | 432 (444.8)[55] | 5.54 (5.5)[56] |
| GaN | 4.51 (4.55)[29] | 199 (203.7)[57] | 3.4 (3.2)[58] |

Table 1. Bulk structural and electronic properties of diamond and zb GaN, where $a_0, B_0,$ and $E_g$ are the lattice constant, bulk modulus, and band gap, respectively. Values in parenthesis are experimental values.

### 3.2. Surface properties

Size effects were studied by varying the surface slab model from one unit cell to three unit cells; slabs have been used to accurately model quasiparticle energies in (111) GaN supercells[20]. The size effects on surface states can be seen in the total DOS for the surface slab, as in Figure 3 for diamond, type I (111) GaN and type II (111) GaN, and Figure S3 for all other GaN surface terminations. Pseudo-hydrogen passivation and dipole correction effects are also shown in Figure 3.

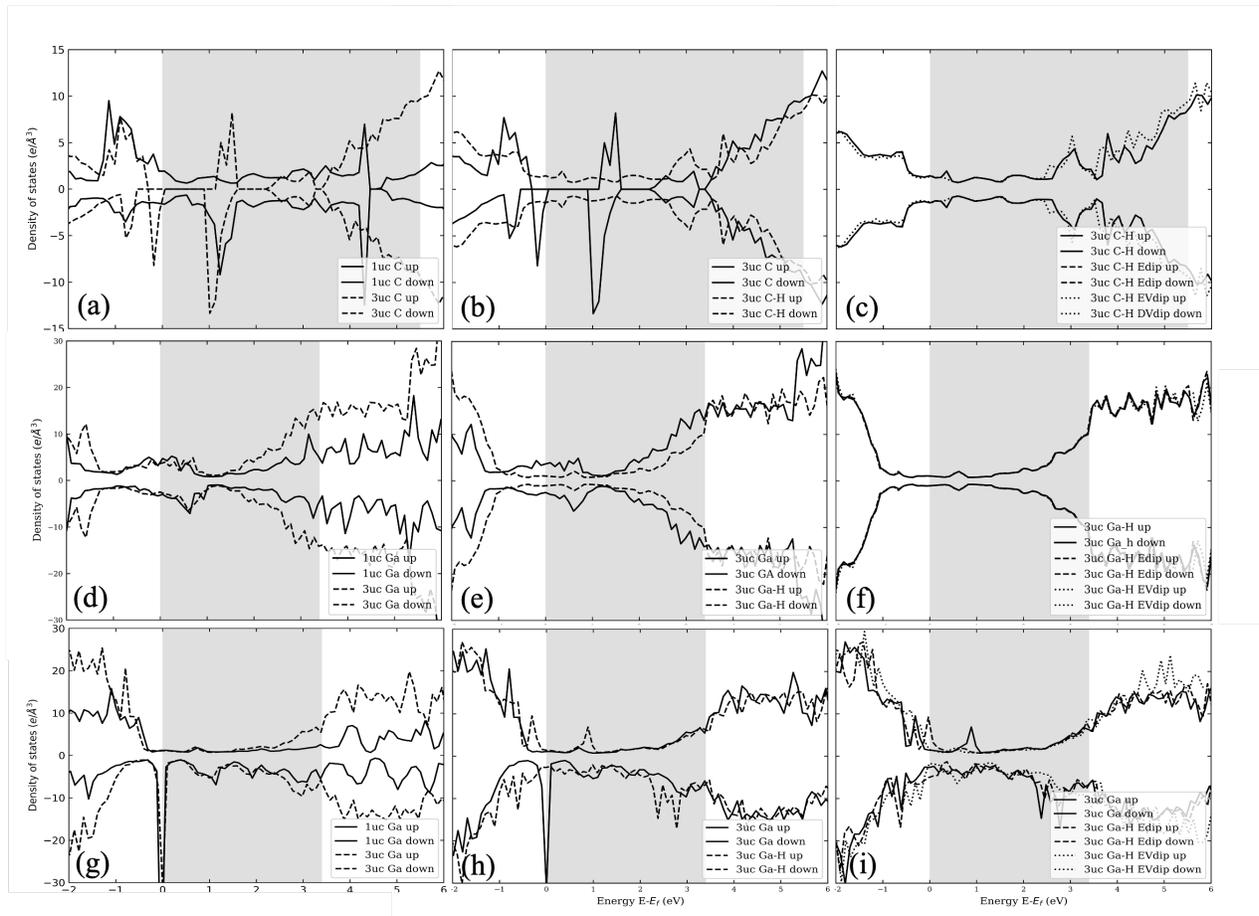

Figure 3. Total DOS (TDOS) for (111) diamond, type I and type II (111) zb GaN. The top row (a-c) depicts diamond, the second row (d-f) depicts type I (111) GaN and the third row (g-i) depicts type II (11$\bar{1}$) GaN. The first column

(a,d,g) compares one unit cell (solid line) and three unit cell (dashed line) supercell TDOS without H passivation or dipole corrections, the second column (b,e,h) compares three unit cell supercell TDOS with (dashed line) and without (solid line) H passivation both without dipole corrections, and the third column (c,f,i) compares three unit cell supercells with H passivation without dipole corrections (solid line), with dipole corrections added to the energy (E) (dashed line) and with total dipole corrections (EV) (dotted line). The grey box indicates the bulk band gap for each material.

Increasing the size of the diamond supercell (Figure 3a) significantly reduces mid-gap states, opening small band gaps near the Fermi energy and near the middle of the band gap; however, the remaining states in the band gap become more localized and a spin-up channel state emerges. Pseudo-hydrogen passivation (Figure 3b) delocalizes these band gap states, removing the gaps due to size effects. This is indicative of the effect of dangling bonds at the surface on band gap states as opposed to the unphysical states present due to periodic image interactions. Dipole corrections (Figure 3c) only have a slight effect on states near the conduction band bottom, increasing and red shifting the conduction band edge states. This negligible change occurs since the polar surface of diamond exists due to locally broken inversion symmetry and not the added effect of electronegativity differences between the constituent atoms as in GaN.

Increasing the size of type I (111) GaN (Figure 3d) blueshifts the near-Fermi level energies slightly and increases the near conduction band states significantly but does not diminish any band gap states. Passivating the bottom Ga surface (Figure 3e) reduces band gap states throughout the gap. Much like in diamond, type I (111) GaN shows negligible changes in the TDOS due to dipole corrections (Figure 3f). As type I (111) GaN is effectively type II (11$\bar{1}$) with a Ga adlayer, this adlayer continues to contribute to the near conduction band edges when the lower surface is passivated; the gap states close to the top of the valence band due to interactions with sub-surface layer N atoms are nearly removed, however due to passivation and dipole corrections.

For the type II (111) GaN surface, increasing the supercell size (Figure 3g) shows negligible effect, slightly increasing the near-conduction band edge states. Pseudo-hydrogen passivation (Figure 3h) reduces the large near-Fermi level state, resulting in a DOS similar to type I (111) DOS; different to type I (111) GaN, however, type II (111) GaN passivated with H reveals semi-local states in the band gap. Unlike diamond and type I (111) GaN however, dipole corrections (Figure 3i) reduce the band gap states created by H passivation. The large near-Fermi level DOS in type II (111) GaN appears to be an artifact of interactions between periodic images, something which is not removed by increasing the size of the supercell but is almost completely removed by passivating the bottom surface and including dipole corrections to remove this unphysical interaction. The systems with passivation and full dipole corrections agree with previous results which used linear combination of atomic orbital methods[59].

Layer dependent DOS can also be used to see the effect of H-passivation and dipole corrections on the DOS as is shown in Figure 4.

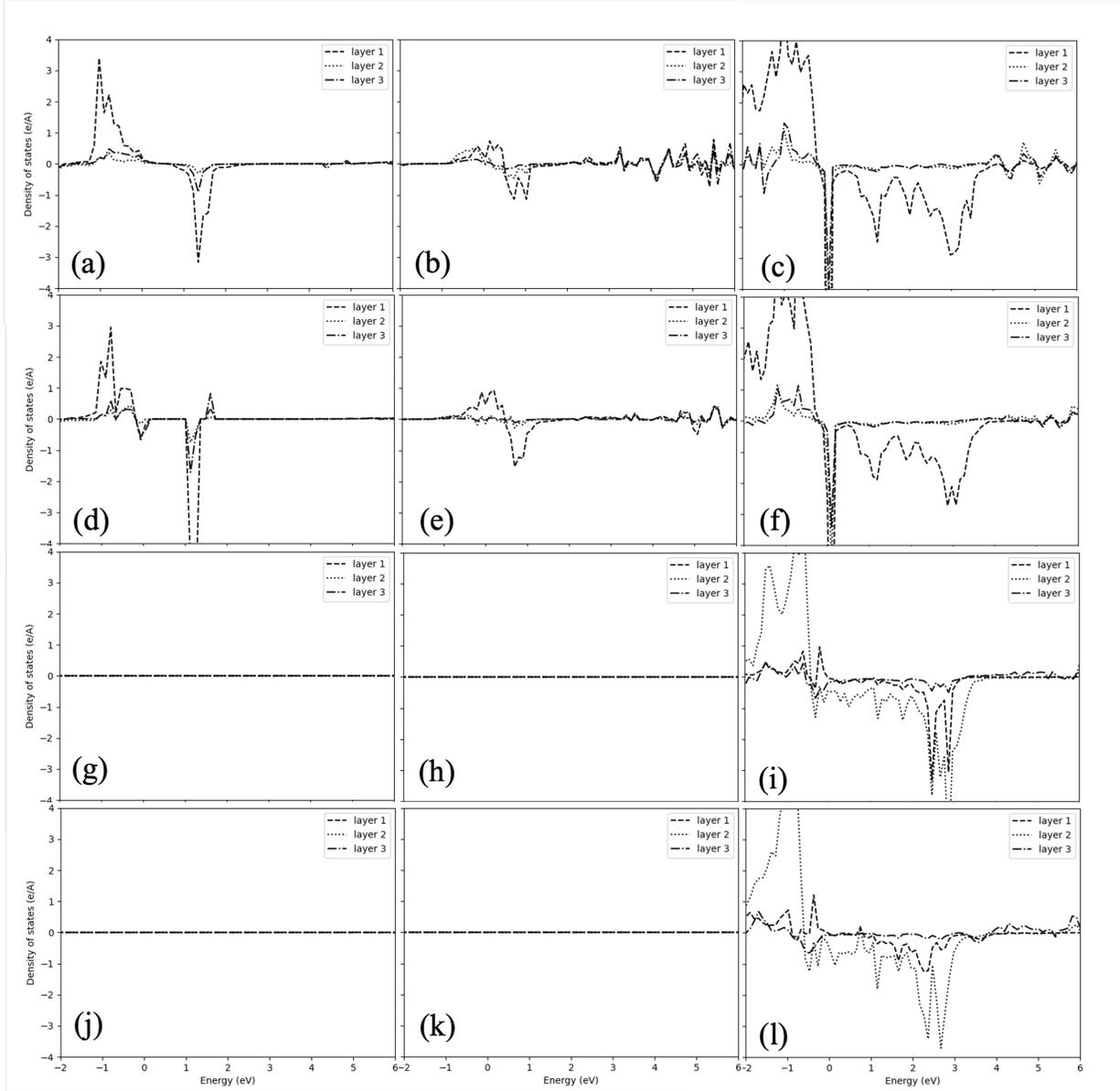

Figure 4. Top layer (dashed line) and two sub-surface layer (dotted and dash-dotted lines) DOS for (111) diamond, and zb type I and II (111) GaN. The first row (a-c) shows three unit cell structures with no H passivation or dipole corrections, the second row (d-f) shows three unit cell structures with H passivation and no dipole corrections, the third row (g-i) shows three unit cell structures with H passivation and energy dipole corrections, and the fourth row (j-l) shows three unit cell structures with H passivation and full dipole corrections. The first column (a,d,g,i) represents diamond, the second column (b,e,h,k) type I (111) GaN, and the third column (c,f,I,l) type II (111) GaN. In all images, the Fermi level is taken to be 0 eV.

The mid-gap DOS state in the (111) diamond supercell without pseudo-hydrogen passivation or dipole corrections is not suppressed by passivation. In fact, this state localizes more and an asymmetry in the spin density emerges. All unphysical surface states are removed when passivation and dipole corrections are added, showing these states were most likely due to interactions between periodic images without dipole corrections to account for local inversion symmetry breaking. With type I (111) GaN, near-Fermi level DOS delocalizes, and the majority of band gap states diminish overall when passivation is added. As with diamond, type I (111) GaN band gap states are completely removed from the top three surface layers when dipole corrections are added, again implying the periodic image interactions persist when only passivation is added. For type II (111) GaN, near conduction band states are reduced while near valence band states persist when the bottom N layer is passivated with pseudo-hydrogen. Interestingly, dipole corrections to the energy localize deep band gap states on the top Ga surface while partially reducing the states on the second N layer and third Ga layer. Also, total dipole corrections including the potential corrections reduce the band gap DOS states for the Ga layers, only shifting the N layer DOS but not necessarily reducing the states.

LPOT plots projected along the surface normal were used to show the conclusive effects of dipole corrections. Figure 5 shows (111) diamond, type I and type II (111) GaN LPOT plots showing size, pseudo-hydrogen passivation and dipole dependence.

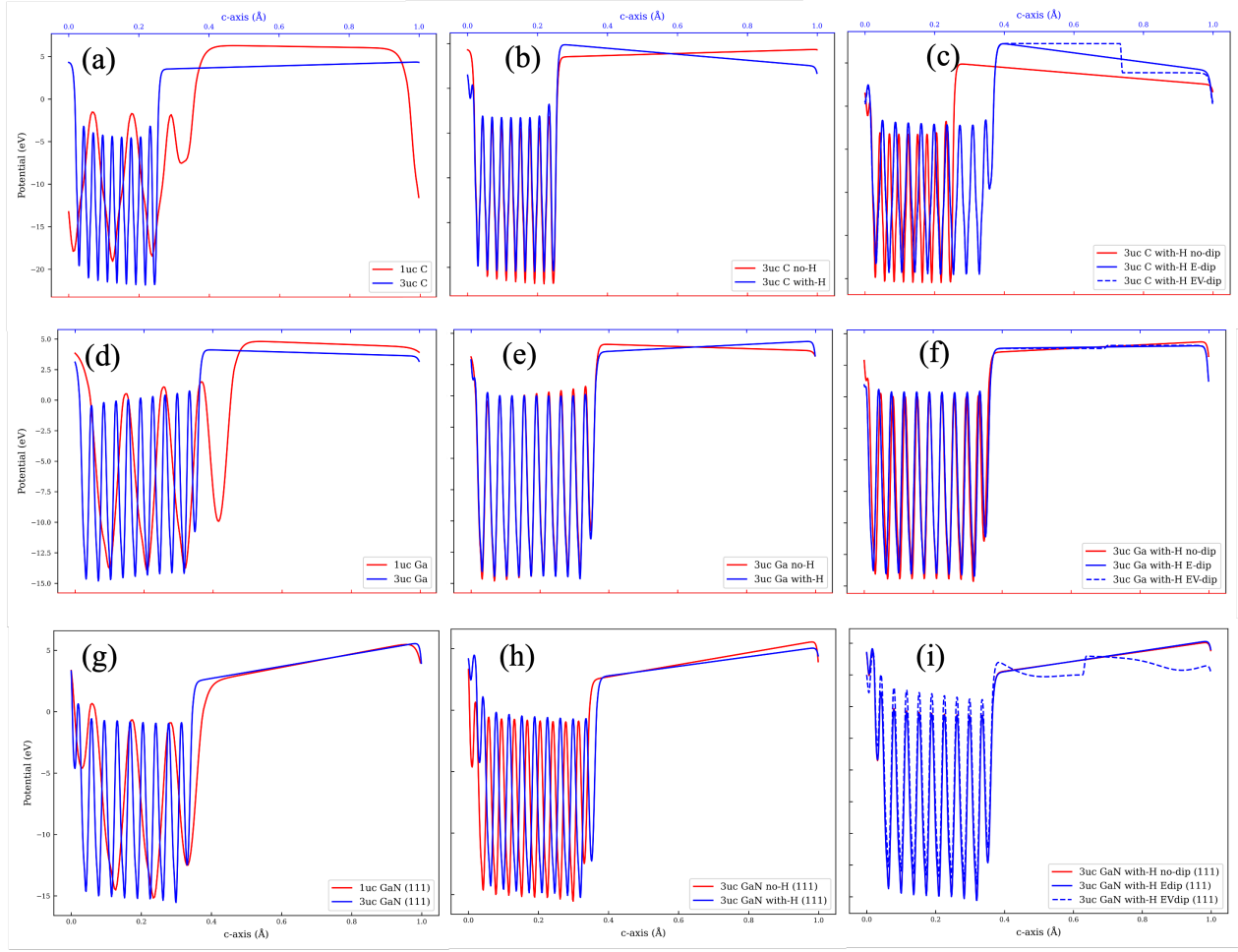

Figure 5. Local potential plots projected along the direction normal to (111) for (111) diamond and type I and type II zb (111) GaN. The first row (a-c) shows (111) diamond, the second row (d-f) shows type I (111) GaN, and the third row (g-i) shows type II (111) GaN. The first column (a,d,g) shows size comparisons between one (red line) and three (blue line) unit cells without H passivation or dipole corrections, the second column (b,e,h) shows H passivation dependence comparisons for three unit cell supercells, and the third column (c,f,i) shows the comparison between three unit cell supercells with H passivation and no dipole corrections (red line), with energy (E) dipole corrections (blue solid line) and with total (EV) dipole corrections (blue dashed line).

In (111) diamond, the LPOT average potential decreases as the supercell size increases and the vacuum potential gradient changes signs (from negative to positive). Pseudo-hydrogen passivation on the bottom C layer increases the average potential and changes the sign of the vacuum potential gradient back to negative. Dipole corrections to the energy cause atomic layers to shift outward while the potential dipole correction removes the electric field at the surface with the typical discontinuity centered in the vacuum region. These corrections show the homopolar nature of diamond where size effects reveal local inversion symmetry breaking, but larger passivated, and dipole corrected supercells show a constant vacuum potential.

The type I (111) GaN LPOT average potential, like in diamond decreases with increasing supercell size, but the vacuum gradient sign is unchanged. Pseudo-hydrogen passivation has little effect on the three unit supercell, however, the sign of the vacuum potential gradient does change. Dipole corrections both to the energy and potential also have negligible effects on the three unit supercell, only removing the gradient in the vacuum potential (along with the central vacuum potential discontinuity).

As with (111) diamond and type I (111) GaN, the average potential in type II (111) GaN decreases with increasing supercell size, while like in type I (111) GaN, the sign of the vacuum potential remains unchanged. Pseudo-hydrogen passivation results in a slight increase in the average potential but, unlike the other two systems, does not change the sign of the vacuum potential. Also, unlike the others, the dipole corrections do not result in a constant vacuum potential. The inclusion of these corrections does not work to remove the interaction between the top and bottom surface in type II (111) GaN, showing that the Ga adlayer (type I (111) GaN) is important for surface stability in zb GaN, as seen in previous studies[20].

Type I (111) GaN shows the most stability with a near constant vacuum potential when looking at all dependences (size, passivation, and dipole correction). Pseudo-hydrogen passivation changes the sign of the vacuum potential gradient for (111) diamond and type I (111) GaN, while dipole corrections remove the electric field at these surfaces along with the expected jump discontinuities in the center of each vacuum region; the periodic image interaction is not fully removed in type II (111) GaN including these corrections.

It is known that materials with different lattice parameters relax such that the lattice deformation minimizes strain in the interface layers[60], which is true for the heterostructures under study as well. Initial relaxation of the three unit cell supercell has a greater effect on GaN surfaces, where the surface layers shift on average by up to an order of magnitude more than in diamond.

Finally, in terms of the surface calculations, the band edge projected charge density may be used to qualitatively study the effects of pseudo-hydrogen passivation and dipole corrections on electronic properties. Figures 6, 7 and 8 show the (111) diamond, type I and type II (111) GaN band edge charge densities, respectively.

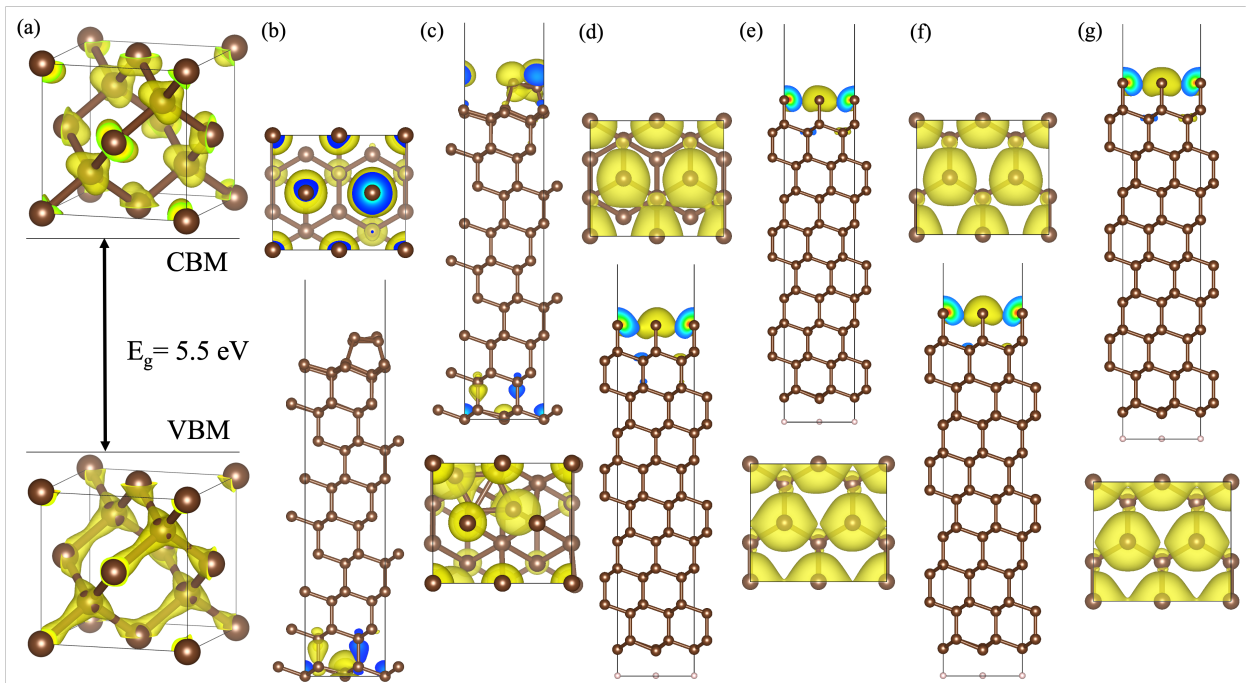

Figure 6. Band edge projected charge density mapping for diamond. (a) Bulk, (b-c) 3-unit cell supercell without H passivation and no dipole correction, (d-e) 3-unit cell supercell with H passivation and no dipole correction, and (f-g)

3-unit cell supercell with H passivation and full dipole correction are shown with a full structure view (side view) and a top-down view of the charge density (looking down the z-axis). The top (bottom) plot in (a) shows the lowest unoccupied orbitals (highest occupied orbitals) of the bulk structure. (b,d,f) show the highest occupied orbital contributions while (c,e,g) show the lowest unoccupied orbital contribution for each system. All isosurfaces are set to 0.2 e/Å$^3$.

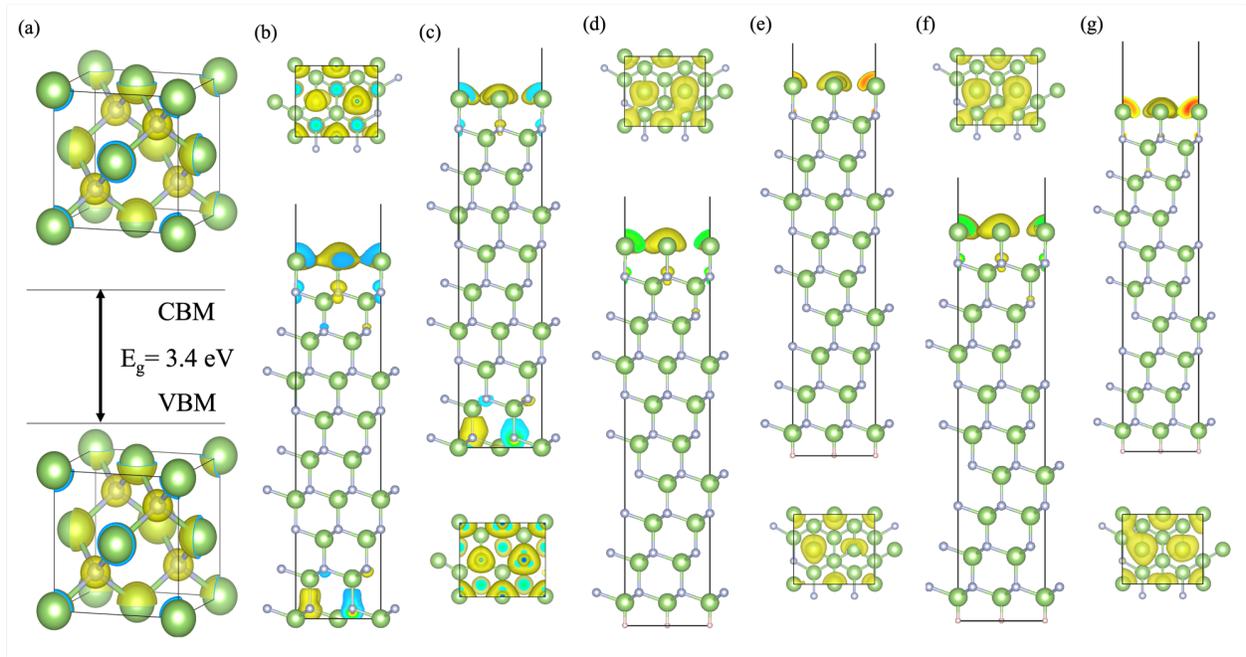

Figure 7. Band edge projected charge density mapping for zb type I (111) GaN. (a) Bulk, (b-c) 3-unit cell supercell without H passivation and no dipole correction, (d-e) 3-unit cell supercell with H passivation and no dipole correction, and (f-g) 3-unit cell supercell with H passivation and full dipole correction are shown with a full structure view (side view) and a top down view of the charge density (looking down the z-axis). The top (bottom) plot in (a) shows the lowest unoccupied orbitals (highest occupied orbitals) of the bulk zb structure. (b,d,f) show the highest occupied orbital (HOMO) contributions while (c,e,g) show the lowest unoccupied orbital (LUMO) contribution for each system. All isosurfaces are set to 0.2 e/Å$^3$.

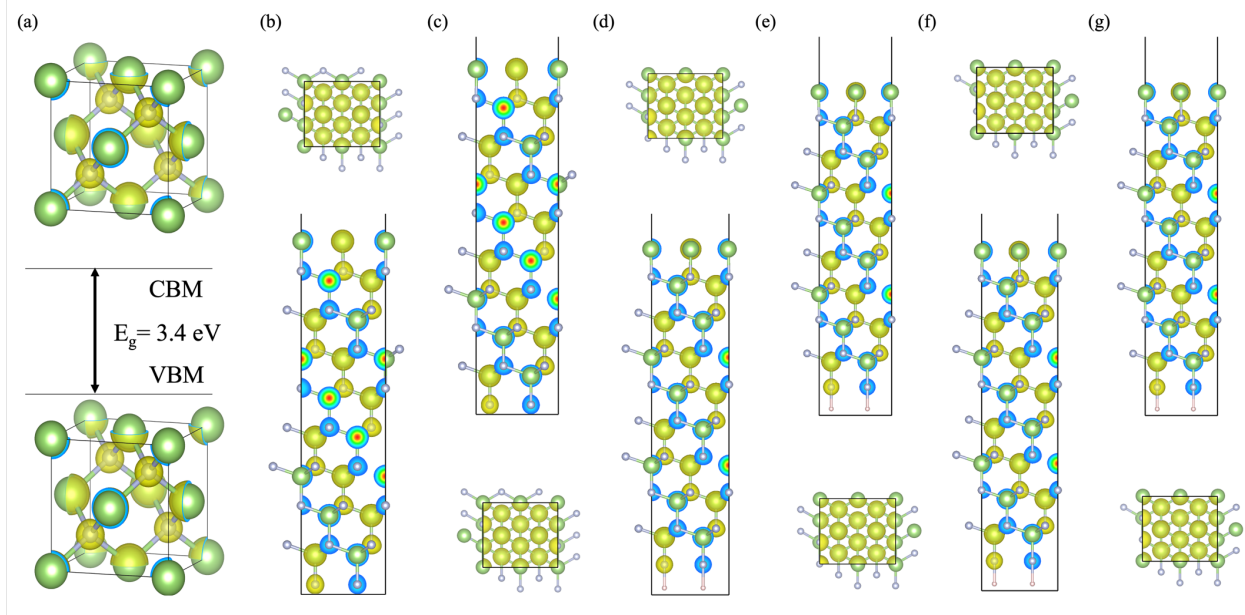

Figure 8. Band edge projected charge density mapping for zb type II (111) GaN. (a) Bulk, (b-c) 3-unit cell supercell without H passivation and no dipole correction, (d-e) 3-unit cell supercell with H passivation and no dipole correction, and (f-g) 3-unit cell supercell with H passivation and full dipole correction are shown with a full structure view (side view) and a top down view of the charge density (looking down the z-axis). The top (bottom) plot in (a) shows the lowest unoccupied orbitals (highest occupied orbitals) of the bulk zb structure. (b,d,f) show the highest occupied orbital (HOMO) contributions while (c,e,g) show the lowest unoccupied orbital (LUMO) contribution for each system. All isosurfaces are set to 0.2 e/Å$^3$.

The effects of H passivation are essentially the same (qualitatively) for diamond and type I (111) GaN surfaces in that passivation removes the band edge charge density from the bottom surface; pseudo-hydrogen works successfully to passivate the bottom layers. Dimerization between top surface layer C atoms in diamond (residing in the C-p LUMO levels) is also removed with H passivation, restoring the correct bulk sp$^3$ bonding throughout the supercell. While the charge density is localized at the surface in passivated, dipole corrected type I (111) GaN supercells, type II (111) GaN charge density is still delocalized across the GaN supercell.

### 3.3. Interface properties

The interface of interest, (111) GaN on (111) C is shown in Figure S6 for type I GaN and type II GaN, along with the relaxed structures without and with dipole corrections added. The average interface layer shift during initial relaxation for the 12 layers surrounding the interface in type I (111) GaN is -1.43 Å and during dipole relaxation is -1.54 Å; a value less than 0 indicates that the layers shift downward towards diamond. The average shift of the diamond layers during initial (dipole) relaxation was -1.35 Å (-1.53 Å) and -1.52 Å (-1.54 Å) in GaN, showing that the system shifts towards the diamond during relaxation. This reveals that zb GaN templates atop and relaxes towards the diamond to relieve the stress due to the large lattice constant mismatch. In the interface between diamond and type II (111) GaN, initial relaxation sees a ubiquitous shift of interface layers upward, away from the diamond; this occurs in both diamond and GaN. On average, layers shifted 2.33 Å during initial relaxation which equates to an average shift of 2.10 Å and 2.55 Å in diamond and GaN, respectively. When dipole corrections were added, the layers all shift back towards the initial atomic positions with an average shift equal and opposite that of initial relaxation (-2.33 Å). Dipole corrections lead to an average shift of -2.29 Å and -2.17 Å in diamond and type II (111) GaN, respectively. This indicates instability in the type II interface. The errant interaction between the (111) and (11$\bar{1}$) surfaces in type II GaN results in an electrostatic and steric instability at the interface which leads to variability in the relaxation of interface layers; in other words, the layers shift one direction initially and shift back towards the initial positions when dipole corrections are included.

The interfacial LPD and CDD for both type I and type II interfaces are shown in Figure 9. For the type I interface (Figure 9, left), the CDD is largest at the interface, returning to a bulk-like density within three sub-layers, both in diamond and GaN. The local potential is unstable at the pseudo-hydrogen surfaces but is more stable in the center of each constituent material, with an interfacial potential (built-in potential) offset of $\Delta V = 3.9\ eV$. The LPD and CDD of Type II (111) GaN on (111) C is unstable throughout the material, showing large fluctuations in both diamond and GaN. Neither pseudo-hydrogen passivation nor dipole corrections work to remove the interaction between the top and bottom layers of GaN. This, as is noted above, is often remedied by utilizing wedge shaped supercell models. A potential offset of more than $\Delta V = 100\ eV$ exists between materials in the type II interface, implying that without the Ga adlayer, as in the type I interface, this heterojunction will prevent charge flow without a very large, applied potential.

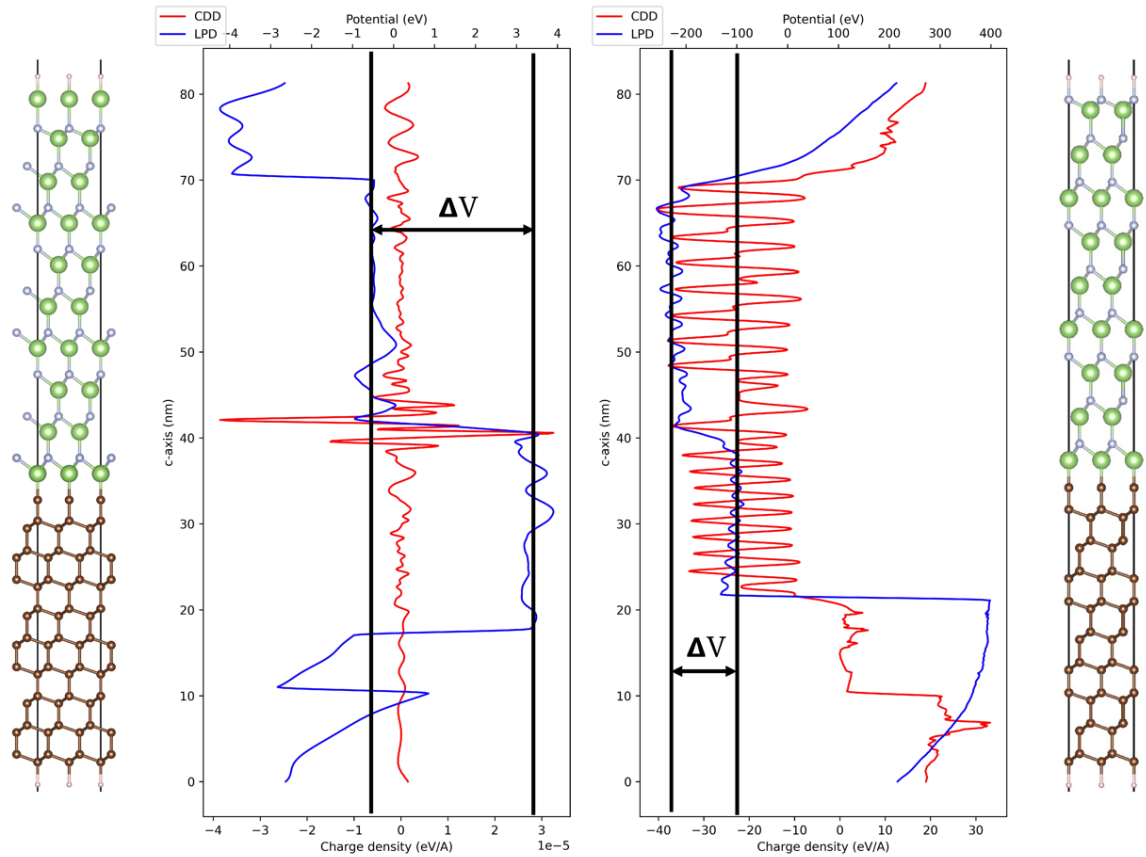

Figure 9. Local potential difference and charge density difference for 3 unit cell type I (111) GaN on (111) diamond (left) and 3 unit cell type II (111) GaN on (111) diamond (right). The interface supercell corresponding to each plot is shown on the sides.

In Figure 10, the layer projected DOS is shown for 12 layers around the interface along with the orbital projected CDD for type I and type II (111) GaN; the bottom six layers are diamond while the top six are GaN. One thing to note about the type I interface (Figure 9, left) is the diamond-like DOS (band gap states) in the first GaN layer (seventh layer from bottom). This implies that the C charge density intercalates into the GaN, something that was seen experimentally using EELS measurements; Ga is not found to permeate into the diamond, while the C signal is non-zero within the GaN interface layers[29]. The DOS in the upper GaN layers then shift towards the conduction band moving further away from the interface. In the type II interface, like in type I, the diamond-like DOS permeates into the first GaN layer. However, the upper GaN layers DOS has now shifted to the valence band edge. This implies that hybridized Ga and N states dominate the upper GaN layers in the type I interface while N-p states dominate in the type II interface.

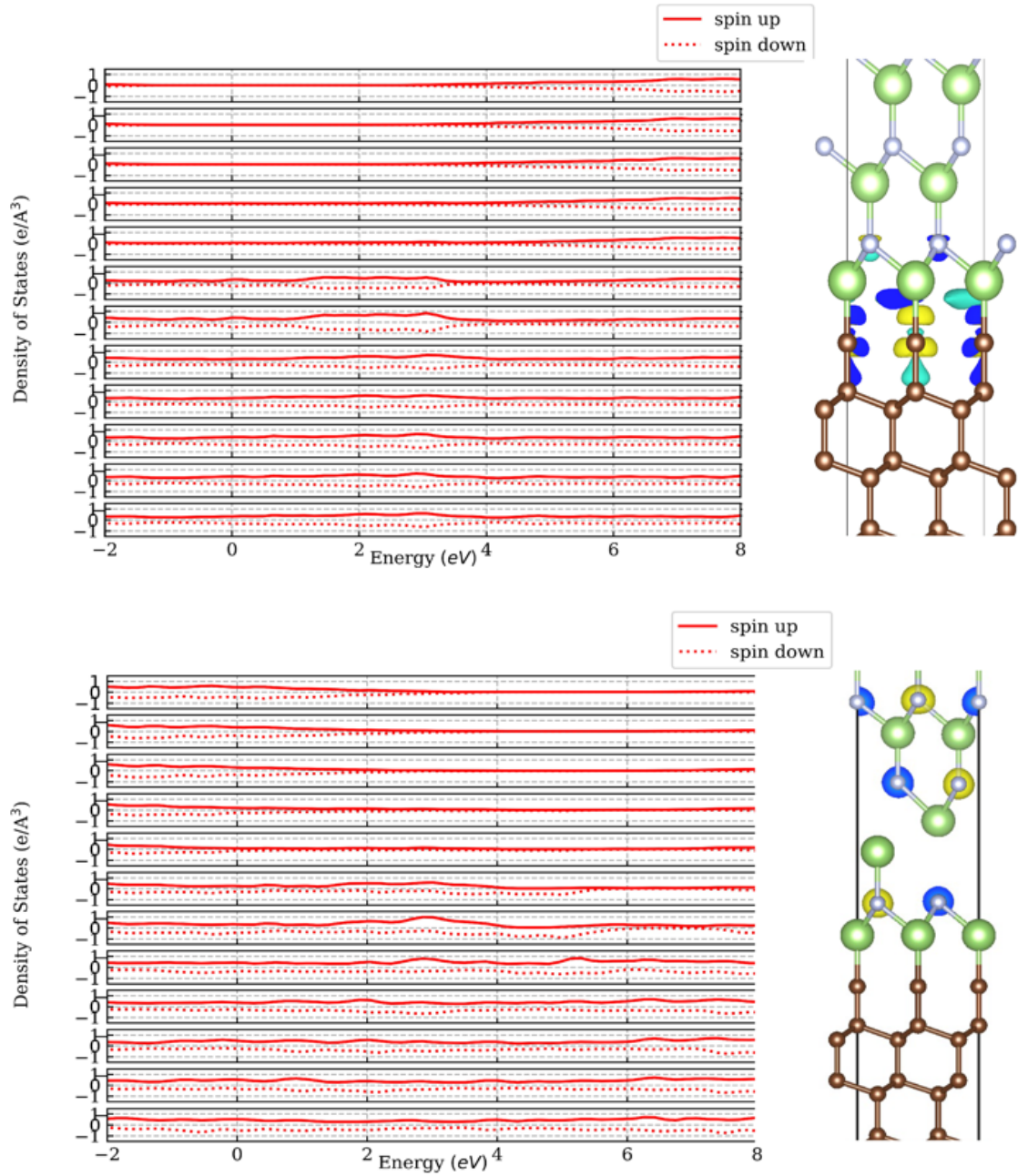

Figure 10. Layer dependent DOS and charge density difference for the 12 interface layers in 3 unit cell type I (upper) and type II (lower) (111) GaN on (111) diamond. The Fermi energy is set to 0 eV.

Interface adhesion energies were calculated using $\gamma = (E_{GaN} - E_C - E_{GaN/C})/A$, where the first two energies terms are the total slab energy of each constituent material, the third energy term is the total interface energy and A is the surface area at the interface. For the type I interface, the adhesion energy is 0.704 eV/Å$^2$ (4.346 J/m$^2$), which is comparable to other III-V zb interfaces[61]. While there exists a relatively sizeable lattice mismatch between zb diamond and zb GaN (~20%), a reasonable reconstruction of the surface atoms combined with the energetics and physical parameters of diamond e.g., bulk modulus, imply this interface can still exist due to templated growth on zb diamond. Thus, in the type I interface, where GaN has a Ga layer at both the top and bottom surfaces, this reconstruction is achieved through the effective addition of a Ga adlayer, which has been shown to reduce stress during growth due to the excess cation layer[62]. For the type II interface the adhesion energy was calculated to be -4.688 eV/Å$^2$ (-75.102 J/m$^2$). As nitrides are known to have high surface energies[63], and a combination of a high surface energy with low interface energy implies stability, the type I (111) GaN on (111) diamond interface is in fact energetically favorable.

## 4. Conclusions

GaN has been used in HEMT application where diamond substrates have shown to improve thermal properties over similar technologies like SiC. Also, lateral overgrowth methods have shown to improve GaN coverage in these devices. Our DFT study has shown that zb type I (111) GaN, effectively a Ga adlayer reconstruction, interfaced with zb (111) diamond is energetically viable due to atomic and electronic reconstruction at the interface. Without this adlayer, however, zb (111) GaN is very unstable. In modeling these polar surfaces and interfaces, type I (111) GaN is stabilized through the inclusion of pseudo-hydrogen passivation and dipole correction added to the energy and potential of the system. These corrections did not result in stabilization of the electronic properties of type II (111) GaN though. Diamond charge density intercalates into the first GaN layer for both type I and type II interfaces, which was also seen experimentally. Therefore, to accurately model optoelectronic properties in polar GaN using slab models, large supercells, pseudo-hydrogen passivation, energy and potential dipole corrections, and a top-layer, cationic adlayer are required.

Finally, as both polymorphs of GaN are present atop a rough diamond surface, wz GaN surfaces and interfaces must also be studied to fully understand the GaN-on-diamond interface. Experimentally, the low energy facet of wz presents as the (0002) surface[29]; thus, the (0001) and (0002) GaN surfaces and interfaces should be studied to clarify why this contracted facet should be the lowest energy. We believe our results help both to clarify methodologies required for modeling polar surfaces and interfaces in GaN/diamond heterostructures and provide a better understanding of the interface present in laterally overgrown GaN/diamond devices. These results also guide future studies on wz GaN/diamond interfaces.

Acknowledgements

The authors would like to thank the Texas State University Learning, Exploration, Analysis, and Processsing (LEAP) HPC cluster for computing resources and for partial support from the Texas State University Research Enhancement Program (REP). We would also like to thank Ed Piner and Jonathan Anderson for helpful conversations and insight.

Supplemental Information

Hybrid density functional theory study on zinc blende GaN and diamond surfaces and interfaces: Effects of size, hydrogen passivation and dipole corrections

1. Bulk modulus calculations

Bulk lattice constants were calculated by varying the value of $a_0$ such that the volume change was < 3%. This ensured enough points were included near the equilibrium volume when fitting the third-order Birch-Murnaghan equation of state. Figure S1 shows the data for bulk diamond and zinc blende (zb) GaN.

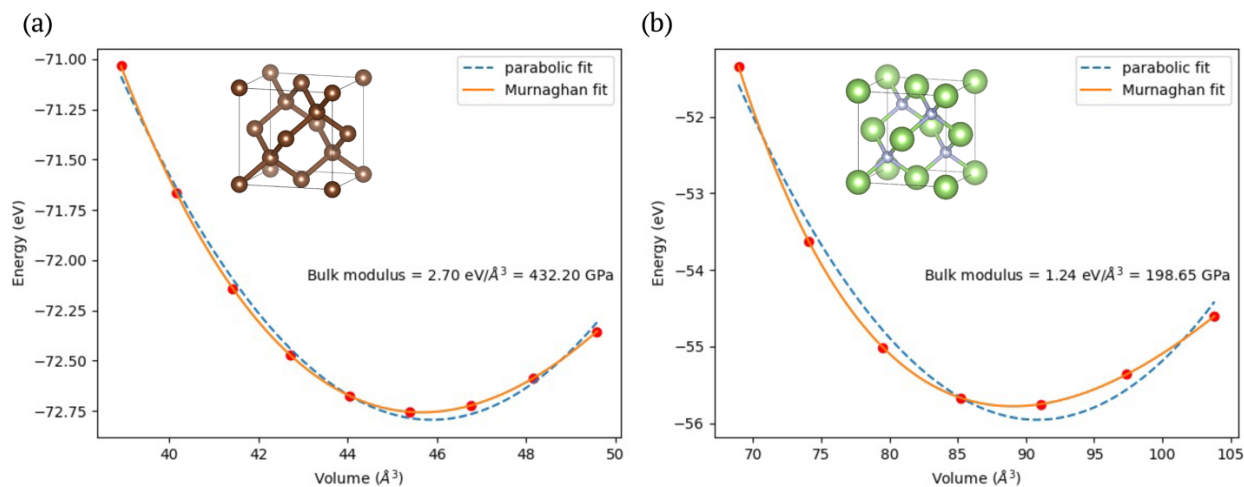

Figure S1. Energy versus volume data with a third order Birch-Murnaghan equation of state (and a parabola for comparison) fit to obtain the bulk modulus for (a) diamond and (b) zb GaN.

2. Surfaces

The optimized bulk structures were then rotated such that the (111) surface was normal to the z-axis using the rotation matrix

$$R = a_0 \begin{bmatrix} -1 & -0.5 & 1 \\ 1 & -0.5 & 1 \\ 0 & 1 & 1 \end{bmatrix}, \quad (S1)$$

where $a_0$ is the optimized lattice constant. Figure S2 shows the three unit cell supercells with pseudo-hydrogen passivation for diamond, type I and type II zb GaN (where the GaN surface types are defined in the main manuscript).

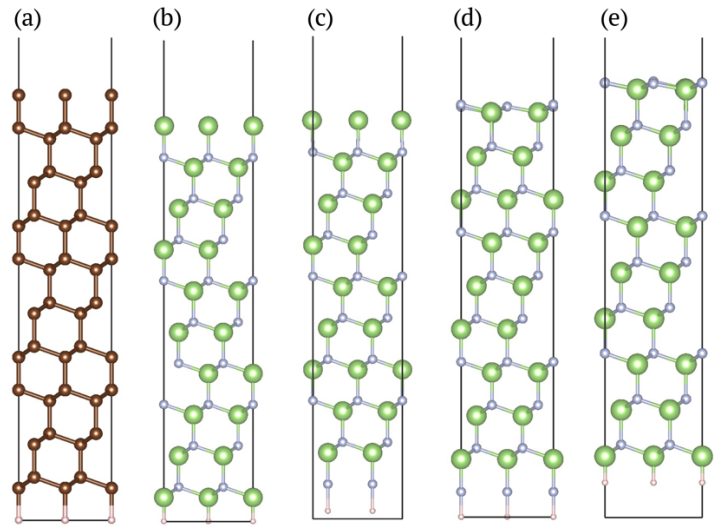

Figure S2. Surface structures for (a) (111) diamond, (b) type I (111) GaN, (c) type II (111) GaN, (d) type I (11$\bar{1}$) GaN, and (e) type II (11$\bar{1}$) GaN. Brown, green, silver, and pink atoms are C, Ga, N, and H, respectively.

Figure S3 shows the total density of states for type I and type II (11$\bar{1}$) GaN. For type I (11$\bar{1}$) GaN, increasing the size of the supercell (going from S3a to S3c) causes the near Fermi level state to localize while the band gap states remain relatively unchanged. Adding pseudo-hydrogen passivation (going from S3c to S3e) results in a slight delocalization of deep band gap states while dipole corrections effect the near Fermi level states. One interesting thing to note is the half-metallic behavior of type I ($11\bar{1}$) GaN, where the spin up channel is semiconducting over most of the band gap (3.4 eV) and the spin down channel is metallic. This has been shown in instances where Ga vacancy defects reveal induced magnetic moments in wurtzite polymorphs of GaN[1] and in similar III-V semiconductors where half-metallicity was seen at the interface between MnSb and GaSb[2].

Type II ($11\bar{1}$) GaN shows a different density of states behavior for the one unit cell supercell (S3b), where the near Fermi level state flips spin; this behavior is also seen for the passivated three unit cell structure without dipole corrections (S3f), while the other plots are similar for type I and type II. Half-metallicity is not seen in type II ($11\bar{1}$) GaN, implying that the adlayer of N atoms is required to preserve half-metallicity. This may be controlled through the flow rate/partial pressure of ammonia during growth.

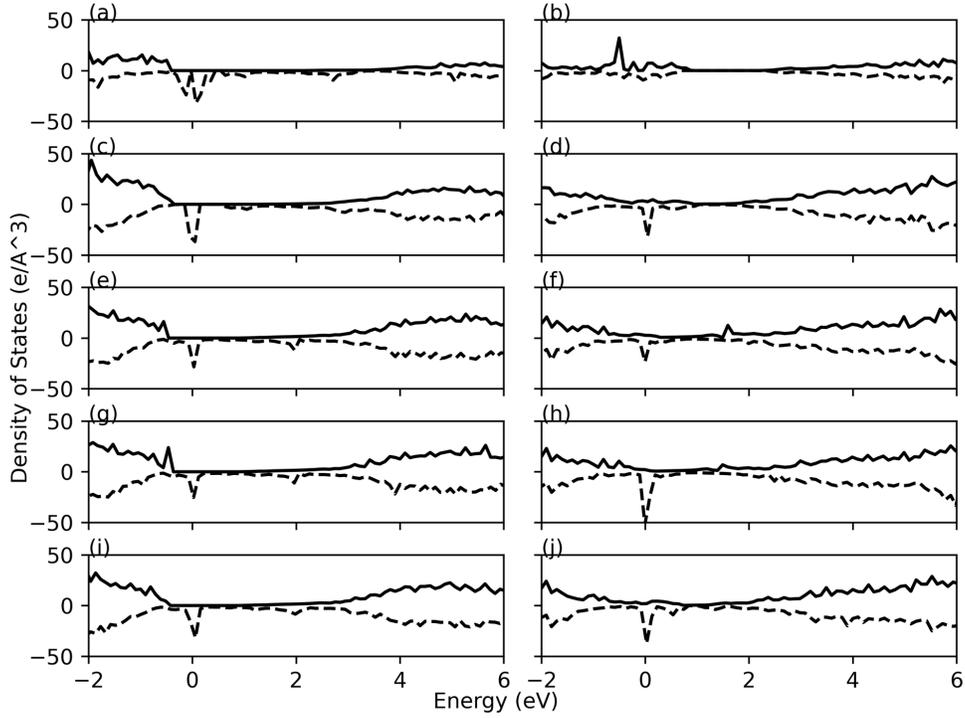

Figure S3. Total density of states for type I (11$\bar{1}$) GaN (first column) and type II (11$\bar{1}$) GaN (second column). The first row shows the one unit cell structures with no passivation or dipole corrections, the second row shows the three unit cell structures with no passivation or dipole corrections, the third row shows the three unit cell structures with passivation but no dipole corrections, the fourth row shows the three unit cell structures with passivation and a dipole correction to the energy, and the fifth row shows the three unit cell structures with passivation and full dipole corrections. The solid line represents the spin up channel while the dashed line represents spin down.

The band edge projected charge density for type I and type II (11$\bar{1}$) GaN is shown in figures S4 and S5 for the bulk and surface structures, where the top surface and side projected structures are shown for surface densities. Both sets of N terminated surfaces show that the ionic charge density remains regardless of the corrections added to the supercell (pseudo-hydrogen passivation and dipole corrections). In type I (11$\bar{1}$) GaN, pseudo-hydrogen passivation removes the negative charge density from the upper surface, but does not localize the charge density at the upper surface like in (111) GaN.

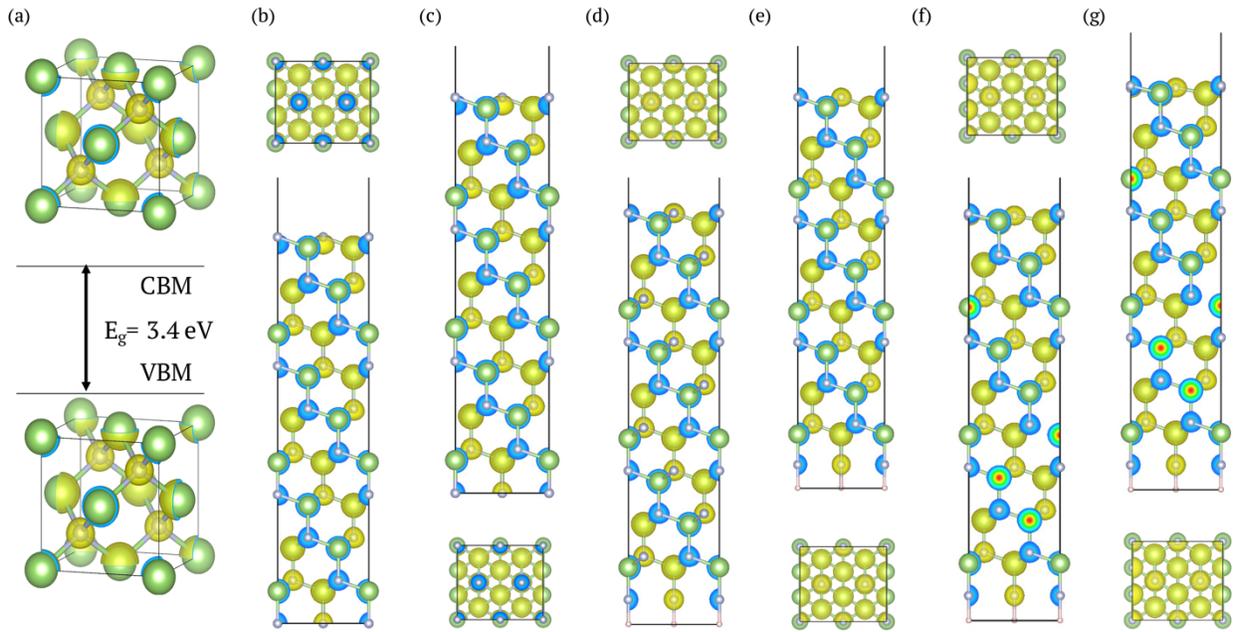

Figure S4. Band edge projected density of states for type I (11$\bar{1}$) GaN. Yellow (blue) isosurfaces represent positive (negative) charge density regions.

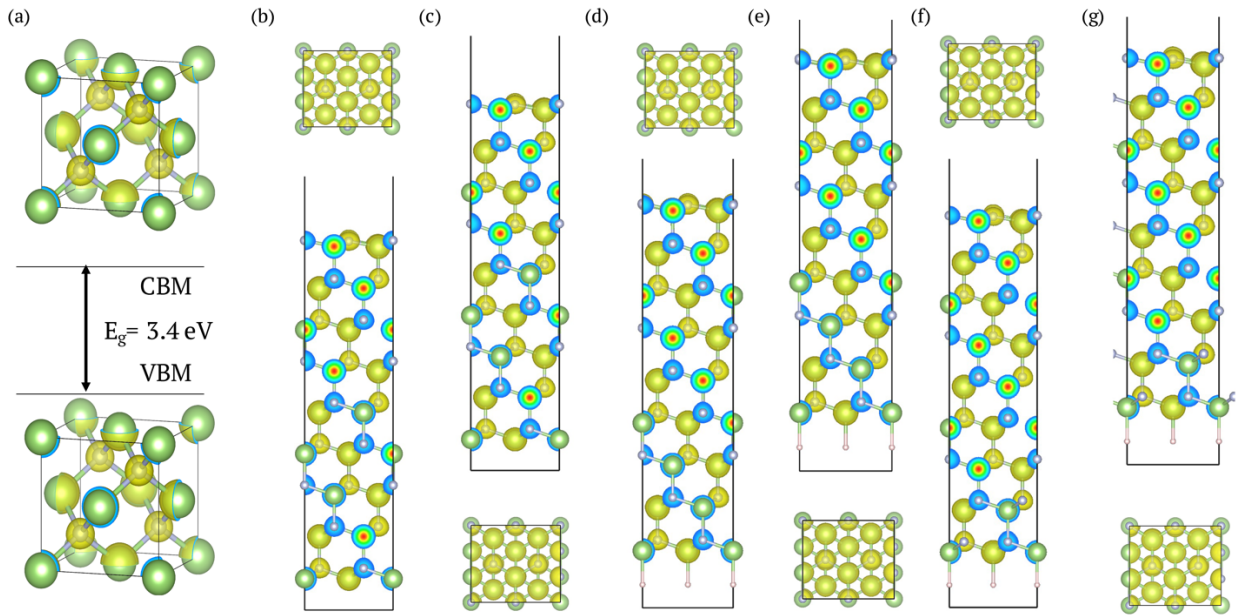

Figure S5. Band edge projected density of states for type II (11$\bar{1}$) GaN. Yellow (blue) isosurfaces represent positive (negative) charge density regions.

The interface structures before and after relaxation are shown in figure S6 for type I and type II $(11\bar{1})$ GaN. Type I $(11\bar{1})$ GaN shifts more at the interface than type II due to the extra layer in type I $(11\bar{1})$ GaN.

3. Interfaces

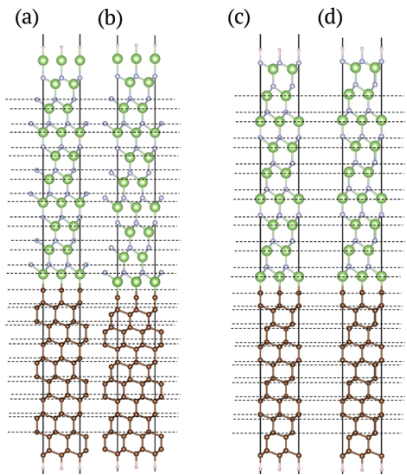

Figure S6. Interfaces for (a,b) type I and (c,d) type II $(11\bar{1})$ GaN showing unrelaxed (a,c) and relaxed structures (b,d). Dashed lines are superimposed to help guide the eye towards the layer shifts after relaxation. Brown, green, silver, and pink spheres are C, Ga, N, and H atoms, respectively.

Half-metallicity in $(11\bar{1})$ GaN is a potential path of interest that is outside the scope of this study; however, this is something which may be exploited and controlled by tunning growth parameters to explore potential device applications. However, the stability of this GaN surface is suspect since the interface energy with diamond is very large as is the built-in potential at the interface. More detailed studies are required to understand the half-metal behavior of $(11\bar{1})$ GaN on diamond interfaces and the potential for applications.